# Millimetre-scale spectrographs for next-generation photonic instrumentation

Emily Ronson*, Robert J. Harris, Ariadna Calcines Rosario, Viktoria Kutnohorsky,
Centre for Advanced Instrumentation (CfAI), Department of Physics, Durham University, South Road, Durham, DH1 3LE, UK; emily.ronson@durham.ac.uk

## ABSTRACT

We present a set of ultra-compact spectrographs fabricated using two-photon polymerisation, each under 1 mm in length and printed directly onto an optical fiber. This fully integrated, alignment-free architecture enables extremely low-mass instrumentation with broad potential across astrophotonics, space systems, drone-based sensing, and quantum optics. Two transmissive configurations are presented; they combine beam-expansion optics, dispersive elements, and stray-light suppression within monolithic printed structures optimised for J- and H-band operation. Coarse gratings provide resolving powers of approximately R ≈ 30. Experimental characterisation confirms the feasibility of fiber-coupled, fully printed spectrographs and demonstrates their promise as deployable components for next-generation integrated photonic instrumentation.

## 1. INTRODUCTION

Spectroscopy is key to modern astronomy. It allows us to probe objects on scales from exoplanets around the stars through the earliest galaxies in the universe. Modern astronomical spectrographs tend to be large instruments, with components that are both costly and time consuming to manufacture. Miniaturized spectroscopy offers significant opportunities across a wide range of markets. Its key advantages include a compact form factor, low weight, reduced thermal mass, and the ability to integrate directly with optical fibers. These characteristics enable applications across multiple fields. In astronomy, this could allow cryogenic instrumentation to be smaller and cheaper. In space and drone-based sensing, the compact and lightweight design reduces payload demands and improves system integration. In quantum sensing, the lower thermal mass compared to current market solutions can deliver substantial operational and cost savings through reduced power consumption and improved thermal control. In biosensing, direct fiber integration creates opportunities for minimally invasive and in vivo sensing applications. [1][2]

Two-Photon Polymerization (2PP) is an additive manufacturing process that uses a tightly focused femtosecond laser to initiate polymerization within a photosensitive resin [3]. Polymerization occurs only at the laser focal point, where two photons are absorbed simultaneously, enabling the creation of highly precise three-dimensional structures with sub-micron feature sizes. By scanning the laser through the resin volume, complex micro- and nano-scale geometries and components can be fabricated layer by layer that would not be possible using conventional manufacturing techniques. In recent years 2PP has been used to produce many optical components, microlenses and freeform lenses are the most common components, the 2PP can be printed directly onto a glass substrate or fibers for easy integration into systems.

The DFG NAIR-APREXIS grant includes a work package to further explore the potential of 3D printing to implement a single compact micro disperser with a coarse transmissive grating (~200lines/mm) to deliver a resolving power at R~10. In this paper we develop two fully integrated fiber coupled photonic spectrographs. In section 2 we introduce optical and mechanical design, the process for manufacturing and the methods for verifying spectrographs. We test their transmission and resolving power presenting the results in section 3, discus the future work of the project in section 4 and conclude in section 5.

# 2. METHOD

## Requirements and initial calculation

For our test spectrograph we aim to develop a micro-spectrograph with a resolving power R~10. We choose the spectrograph to match our laboratory camera high QE range (.9-1.7 µm) working in spectral order 4 (1 – 1.225 µm) and order 3 (1.33 -1.65 µm).

## Optical design

We produced two different optical designs, which are shown in Figure 1. The first is based upon a more traditional design philosophy, with a separate collimating/focussing lens and grism assembly (multi-element design). The second combines the main optical elements (collimator and grism) into one (single element design). We focus on optimising our design for our laboratory detector, a Xenics Xeva, with high QE from .9 to 1.7 µm.

Both designs are fed by a single on axis single mode optical fiber (a P1-980A-FC from Thorlabs), with a numerical aperture (NA) .13-15 and a mode field diameter (MFD) of 5.3 - 6.4 µm (at 980 nm). The beam passes through a spherical initial beam expansion lens with $R_c$ = 25 µm. From here both collimate using an aspherical lens and then slowly focus the light (F~23.5) approximately 3.7 mm away. Both designs use a grism with 0.06 lines/ µm to give a theoretical maximum resolving power between 30-40 assuming 12 rulings will be illuminated. As the input is an on axis single mode fiber, the grating size does not depend on distance to the pupil reducing the length of the spectrograph. We note that the lens directions are reversed in the two designs. On one hand an upward facing lens is simpler to design and print so is suitable for the multi element design, on the other hand it is not possible to integrate this shape into a single element with the grism. Our slow beam is low aberration, with diffraction limited performance across the entire band. An example of this is given in Figure 2.

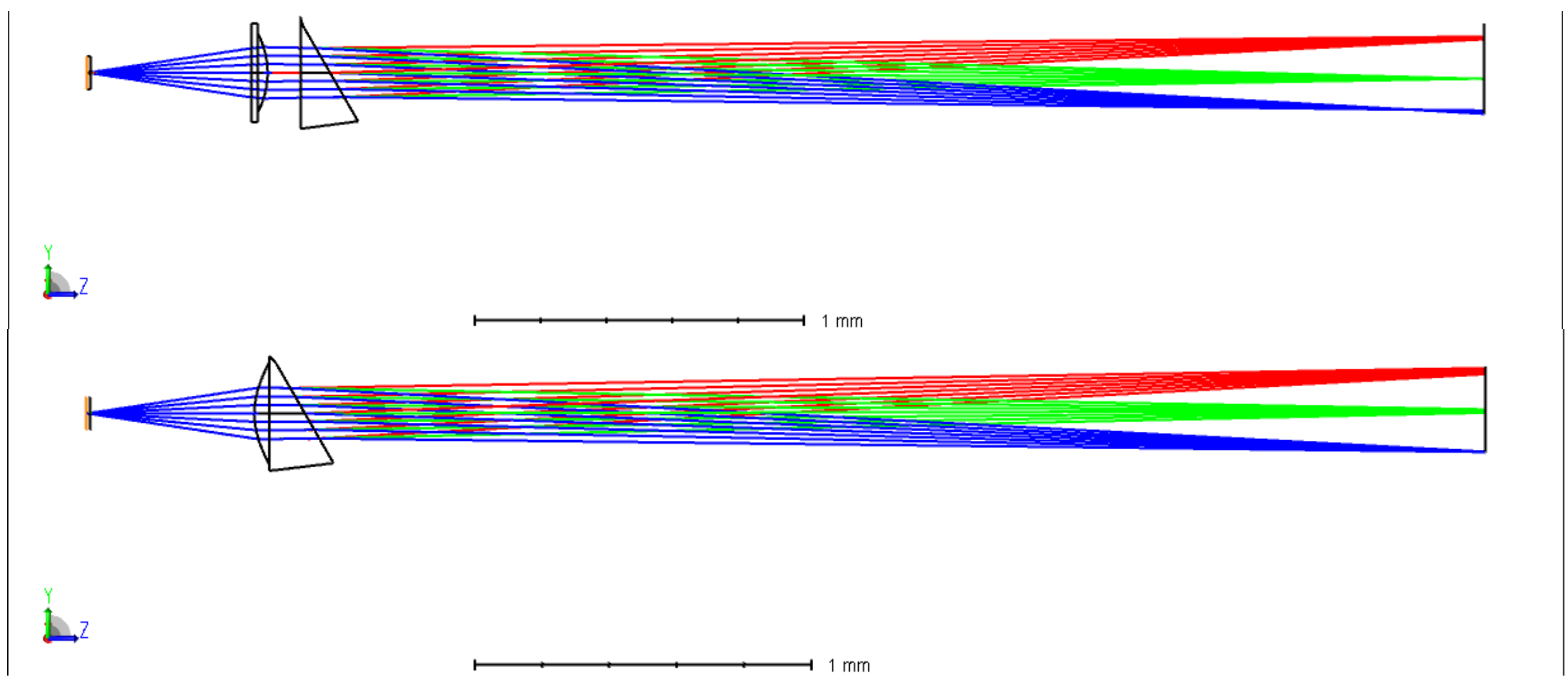

*Figure 1 Top: Multielement Zemax design consists of a beam expander, a collimating lens and a grism. Bottom: Single Element Zemax model, consists of a beam expander and a collimator and grism assembly in one piece.*

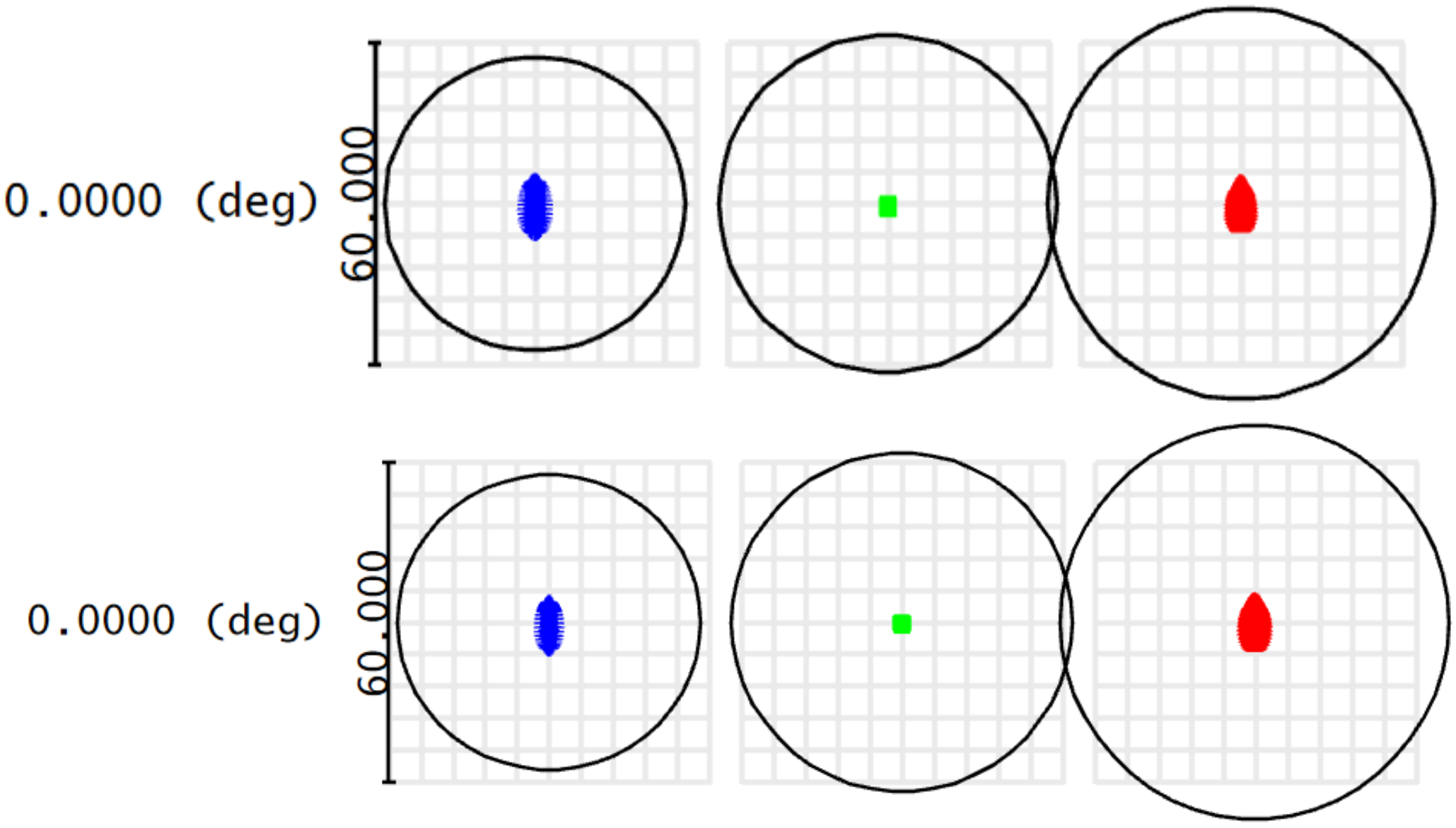

*Figure 2 Top multielement spot diagram showing wavelengths 1, 1.1 and 1.225, and bottom shows the spot diagram for the single element design at the same wavelengths.*

**Mechanical design**

The mechanical design was developed to satisfy the constraints associated with fabrication via two-photon polymerization (2PP). For the selected print configuration, the minimum achievable feature size was approximately 100 nm. Hollow features were designed with dimensions of at least 20 µm to facilitate removal of uncured photoresist. Unsupported beam lengths were limited by the mechanical stability of the printed polymer during fabrication and development; in this work a conservative maximum span of 140 µm was adopted based on empirical process characterization and the known challenges associated with long overhanging structures in 2PP [4].

The mechanical assembly was created by importing the Zemax optical design into CAD and constructing a supporting structure around the optical components. This approach enabled the complete spectrograph assembly to be fabricated as a single monolithic print in one material. A key consideration for the design is managing stray light as the mechanical structure is the same material as the lenses so stray light can propagate inside the mechanical structure. The addition of cavity walls in the design was for the addition of IP-Black. This is a separate post processing step where the cavity was filled with a black material to supress stray light within the spectrograph. The internal walls have been roughened with a sinusoidal wave shape to further reduce stray light. The designs also incorporate enclosed internal cavities, so access tubes have been added to facilitate removal of residual photoresist from these volumes. The CAD model is shown is Figure 3.

As with the optical design the grating is designed to have 0.06 lines/µm which equates to 12 lines across the field of view. The grating has 90-degree steps, with the upper surface parallel to the fiber tip (horizontal) this will result in no additional stepping during the printing process from building up the shape through multiple passes.

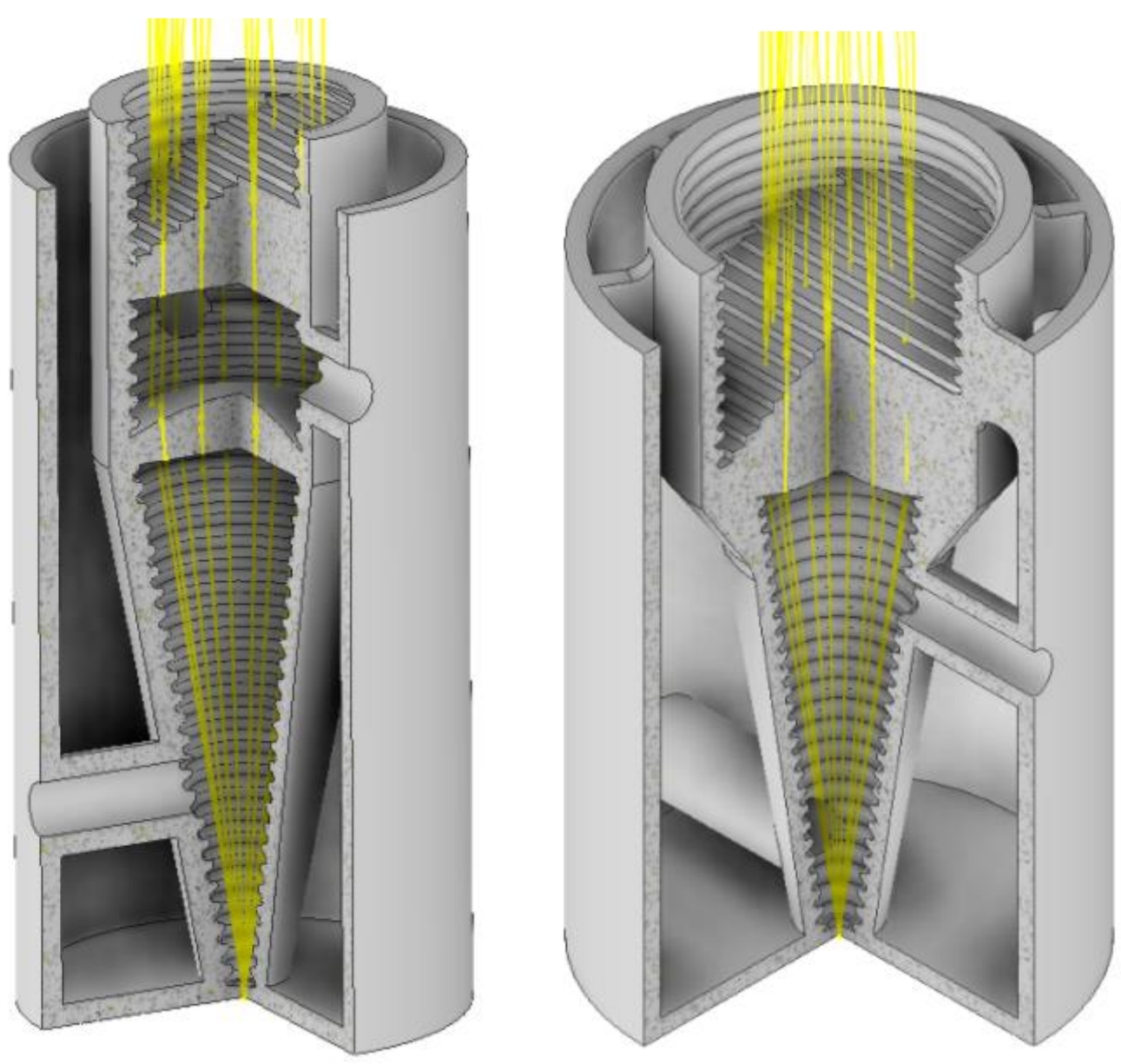

*Figure 3 CAD models of the spectrographs. On the left the multielement design has a height of 815µm and outer diameter 200µm. On the right is the single element design which has a height of 794µm tall and an outer diameter of 240µm.*

## Manufacture

The devices were fabricated by Printoptix with the highest resolution that was offered at the time (63x lens with IP-Dip photo resist). Figure 4 shows the prints before any blackening. IP-Dip2 (Nanoscribe GmbH) was selected as the photoresist for two-photon polymerization due to its high resolution, excellent dimensional stability, and favorable optical properties. The material is specifically formulated for direct laser writing and enables the fabrication of complex three-dimensional microstructures with sub-micron feature sizes. For optical applications, IP-Dip exhibits high transparency across the visible and near-infrared wavelength ranges, low optical absorption, and a refractive index of approximately $n$ = 1.52–1.54 depending on wavelength and processing conditions. [5]. A proprietary black polymer was added as a post processing step. [6]. It was found that the multi-element spectrograph was more difficult to manufacture, due to the complexity of printing multiple optical elements, this will be fed back into future work.

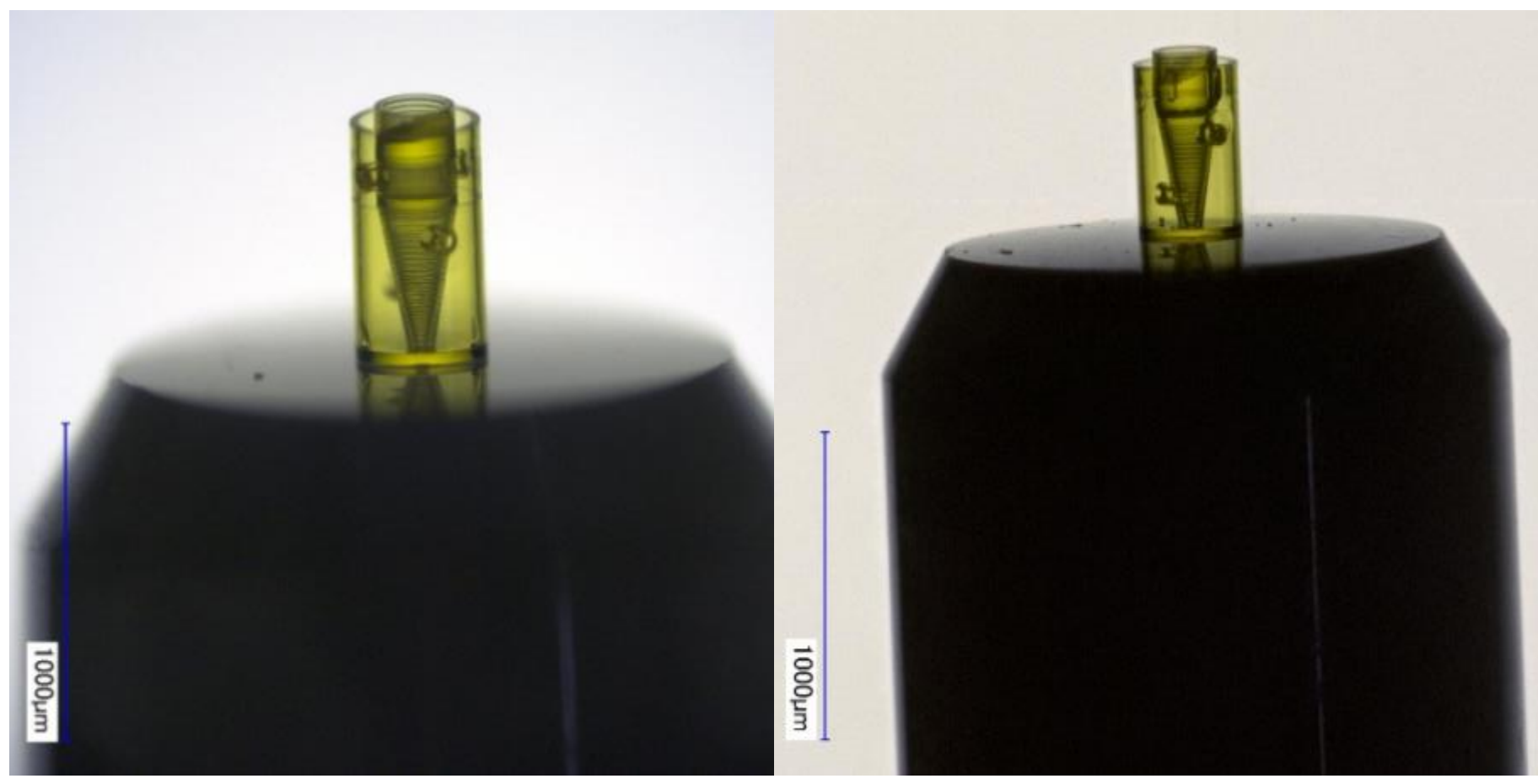

*Figure 4: Images of the printed spectrographs on an FC fiber ferrule before the black polymer is in added. The left shows the multielement design and the right the single element design both are shown printed onto the fiber tip.*

### Experimental verification

Once printed and shipped the two designs were verified in the lab. To characterize the spectrographs two tests were performed to determine the transmission and spectral resolving power. For both tests we used a SuperK Evo EU-4 with a Select IR, both from NKT as a light source. This allowed us to select wavelengths of interest in the 1100-1800 nm band and access sufficient illumination for high signal to noise measurements.

**Test 1 – Transmission of the spectrograph**

To calculate the transmission of the spectrograph the superK fiber was coupled to a test patch cord of the same length and type as the printed spectrograph fiber and then fed into a power meter (Thorlabs PM16-122). Power was measured as a reference measurement for optimum transmission. We then replaced the test patch cord with the printed spectrograph and recorded a measurement on the power meter. As with the resolving power test, the superK wavelength varied from 1100-1800 nm in 100 nm steps.

**Test 2 – Resolving power of the spectrograph**

To test the resolving power, the superK output fiber was coupled to the unprinted end of the spectrograph fiber using a ADAFC2 mating adaptor from Thorlabs. Due to the short focal length of the spectrograph a 4-f re-imaging setup was created using two Thorlabs AC127-050-C lenses and the output was imaged onto a Xenics Xeva-1.7-320. The superK wavelength was varied between 1100 and 1800 nm in 50 nm steps, finally a "flat" to allow spectral tracing was taken using multiple wavelengths at once. A schematic of the test setup is shown in Figure 5.

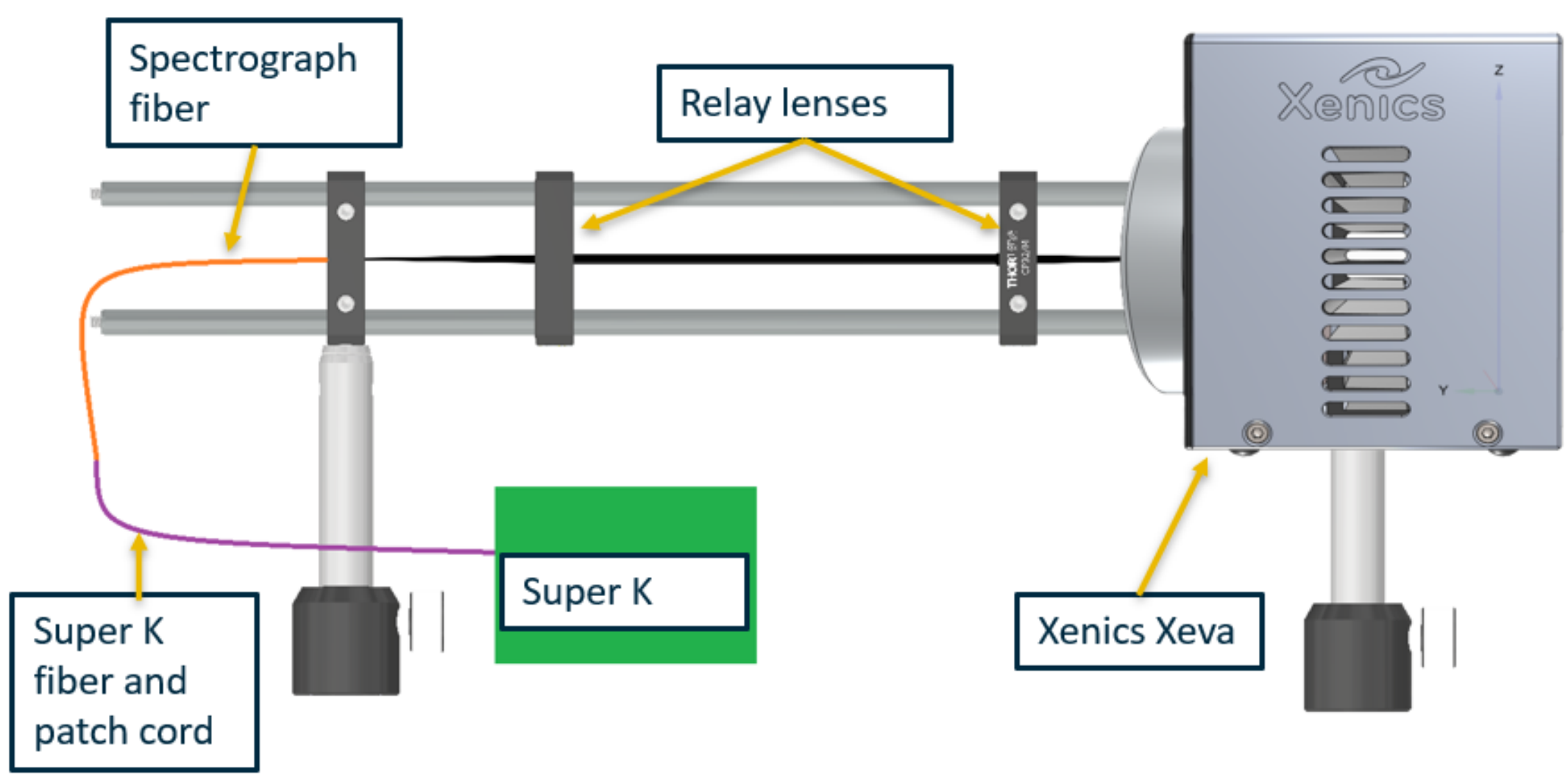

*Figure 5 Test set up for imaging the spectrograph. The test set up includes a SuperK laser with fiber coupled to the spectrograph fiber, a relay lens is used to re-image the focus and an Xenic Xeva-1.7-320 camera to image the output.*

## 3. RESULTS

The data below shows preliminary results for the first tested fibers; future work will characterize the spectrographs in more detail.

**Spectrograph Transmission**

The averaged results of the transmission of the dummy fiber and both spectrographs are shown in Figure 6, both spectrographs as a percentage of transmission are shown in Figure 7. The bare fiber shows the highest power measurement, followed by the single element design and finally the multielement design. All the plots show a sharp transmission decrease after 1600nm, likely due to bend losses in the fiber. Overall, the single element design shows a transmission of ~80% over the range of 1100-1600, and the multielement design between 50 and 60%.

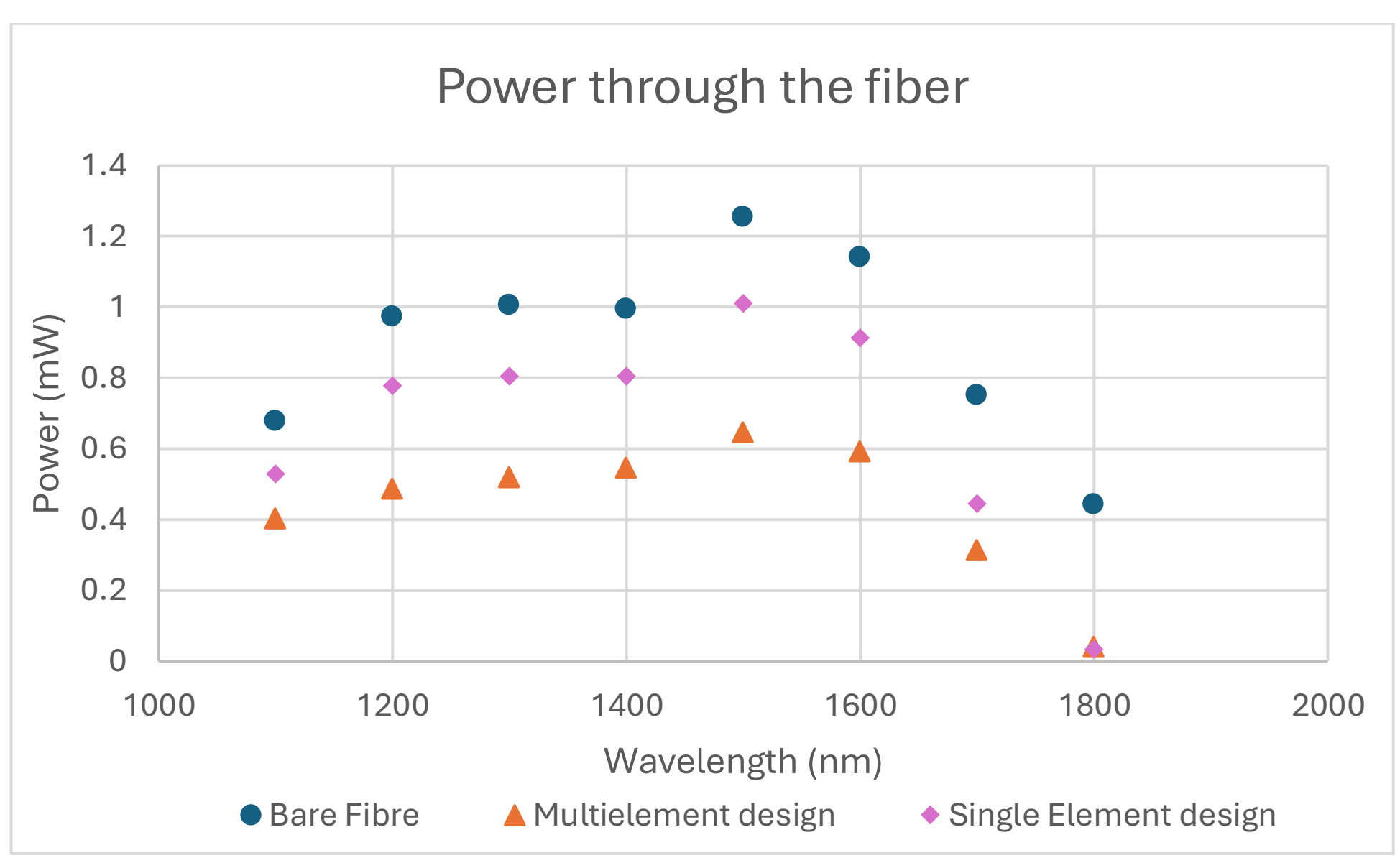

*Figure 6 Transmission results for both spectrograph designs, along with transmission for the bare fiber.*

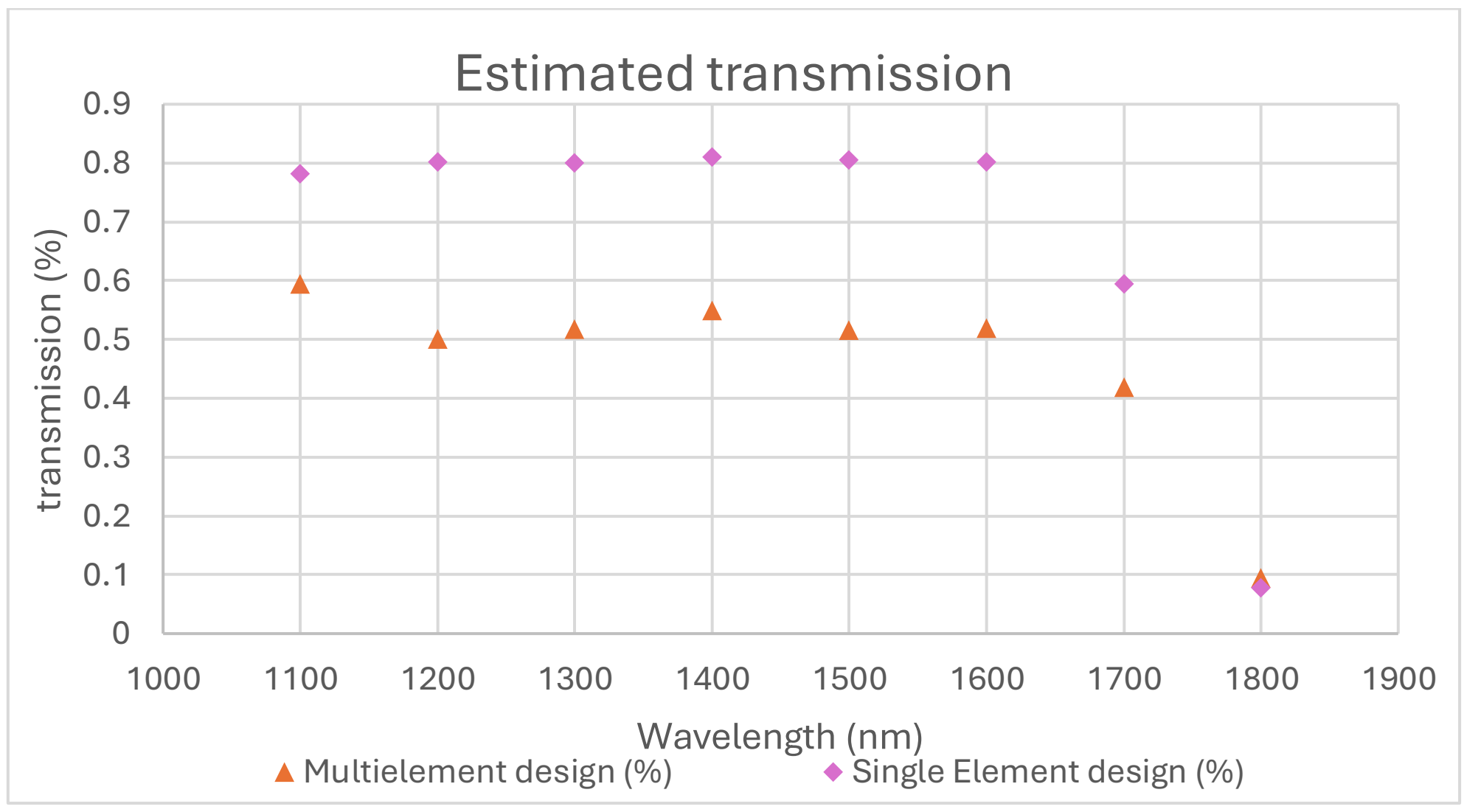

*Figure 7 Estimated percentage transmission through the spectrograph, for the multielement design the average is 53% (excluding data above wavelength 1550nm) and the single element design is 80%.*

**Spectrograph resolving power**

The data from the Xenics Xeva was processed using custom software. The peaks of each illumination wavelength image were fit using a centre of mass model and then a Chebychev polynomial was used to give a wavelength solution. A gaussian was then fitted to each peak and the FWHM used to estimate the resolving power. The resolutions from each wavelength measured at each order are summarised in Table 1 and shown in Figure 8.

*Table 1 Calculated resolutions for both designs for the three orders measured*

| Wavelength | Resolution | |
|---|---|---|
| | Multielement design | Single element design |
| 1100nm – 1200nm (Order 4) | 23.8 ± 0.1 | 26.0 ± 1.2 |
| 1250nm – 1650nm (Order 3) | 21.0 ± 0.4 | 22.0 ± 1.8 |
| 1700nm – 1850nm (Order 2) | 14.6 ± 2.5 | 16.0 ± 2.7 |

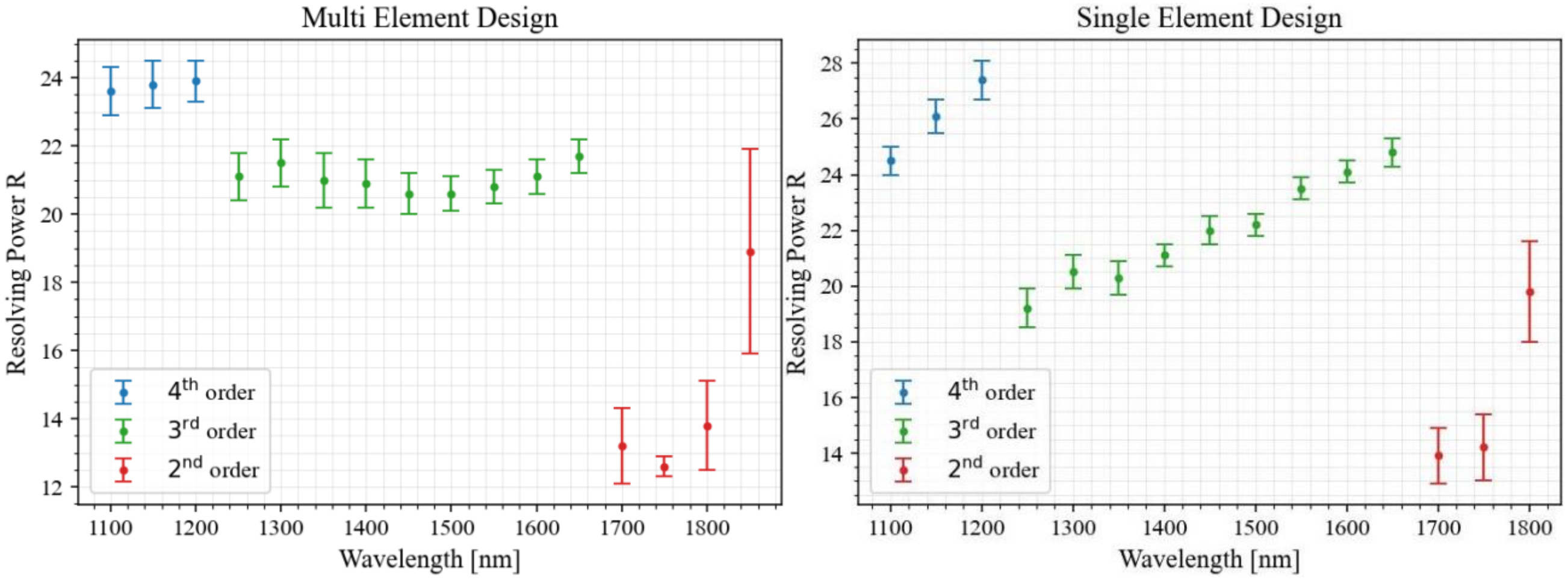

*Figure 8 Spectral resolution for the multielement design (Left) The mean resolution in order 4 is 23.8 ± 0.1 and in order 3 21.0 ± 0.4. Spectral resolution for the single element design (Right) where the mean resolution in order 4 is 26.0 ± 1.2 and in order 3 is 22.0 ± 1.8. The second order (red points) are for reference and not included in the mean resolving power estimates.*

## 4. DISCUSSION

Our results show that fabrication of microscale spectrographs on fiber tips is feasible. The transmission measurements show significant improvement (20-30%) in the single element design, suggesting that combining printed elements is a sensible solution, difficult to achieve with conventional optics.

The resolving power measurements show comparable resolving power for both designs. We find a mean resolving power of ~20, lower than the theoretical value of 30-40. Whilst this is still above our requirement of 10, further investigations are required to determine the cause. The investigations will include tests and analysis at multiple focus positions, a study the proportion of the grating that has been illuminated to determine how many rulings are illuminated and do an in-depth characterization of the spectrograph and images to establish any aberrations affecting the system. Straylight is another important consideration, with future tests we will use a lower noise detector and compare Zemax stray light results to our experimental results.

We note the good performance of the SM980 fiber until 1600 nm, where there is a sharp drop in transmission. This is likely due to bend losses as SM980 is designed for use in the 980-1550 nm wavelength range.

We will also optimize the design for future astronomical testing. This will include grating order optimization to avoid the atmospheric absorption between the astronomical bands.

## 5. CONCLUSION

We present two micro spectrographs created using two photon polymerisation. The spectrographs are < 1 mm and are printed on SM980 single mode fibers. We show a transmission of up to 80% across a wide wavelength range (1000 – 1600 nm) and a spectral resolving power of ~20 across spectral orders 3 and 4. Due to their size and performance, the devices show real promise across fields including astronomy, quantum studies and space science. Further work will include testing multiple versions of each design, investigating multiple focus positions and an in-depth characterization of the structure. We will also begin work on a reflective design using 2PP and sputter coating.

## 6. ACKNOWLEDGEMENTS

We would like to acknowledge our collaboration with Nils Fahbach, Kay Glaser and Simon Thiele from Printoptix who printed the spectrograph samples discussed in this paper. RJH received support as part of the "NAIR-APREXIS" grant, funded by the Deutsche Forschungsgemeinschaft (grant no. 506421303).

## REFERENCES

[1] H. Wang, W. Zhang, D. Ladika, H. Yu, D. Gailevičius, H. Wang, C.-F. Pan, P. N. S. Nair, Y. Ke, T. Mori, J. Y. E. Chan, Q. Ruan, M. Farsari, M. Malinauskas, S. Juodkazis, M. Gu, J. K. W. Yang, Two-Photon Polymerization Lithography for Optics and Photonics: Fundamentals, Materials, Technologies, and Applications. *Adv. Funct. Mater.* 2023, 33, 2214211. **https://doi.org/10.1002/adfm.202214211**

[2] Andrea Toulouse, Johannes Drozella, Simon Thiele, Harald Giessen and Alois Herkommer, 3D-printed miniature spectrometer for the visible range with a 100 × 100 µm2 footprint, Light: Advanced Manufacturing (2021)2:2 https://doi.org/10.37188/lam.2021.002

[3] Linjie Li, John T. Fourkas, "Multiphoton polymerization," Volume 10, Issue 6, Materials Today, ISSN 1369-7021, https://doi.org/10.1016/S1369-7021(07)70130-X (https://www.sciencedirect.com/science/article/pii/S136970210770130X)

[4] Marschner, D.E., Pagliano, S. and Niklaus, F. (2022) A Methodology for Two-Photon Polymerization Micro 3D Printing of Objects with Long Overhanging Structures. SSRN. Available at: https://ssrn.com/abstract=4203217

[5] MICHAEL SCHMID,* DOMINIK LUDESCHER, AND HARALD GIESSEN, "Optical properties of photoresists for femtosecond 3D printing: refractive index, extinction, luminescence-dose dependence, aging, heat treatment and comparison between 1-photon and 2-photon exposure," Vol. 9, Vol. 9, No. 12 / 1 December 2019 / Optical Materials Express, 4564-4577 (2019), https://doi.org/10.1364/OME.9.004564

[6] Michael D. Schmid, Andrea Toulouse, Simon Thiele, Simon Mangold, Alois M. Herkommer, and Harald Giessen "3D Direct Laser Writing of Highly Absorptive Photoresist for Miniature Optical Apertures", 202211159, Advanced Functional Materials, December 2022, DOI:10.1002/adfm.202211159